\documentclass[pra,nofootinbib]{revtex4-1}
\usepackage{graphicx}
\expandafter\let\csname equation*\endcsname\relax
\expandafter\let\csname endequation*\endcsname\relax
\usepackage{amsmath}
\usepackage[]{amssymb}
\usepackage{color}
\usepackage[colorinlistoftodos]{todonotes}
\graphicspath{{}}

\def\bra#1{\left\langle#1\right|}
\def\ket#1{\left|#1\right\rangle}
\usepackage{stackengine}
\def\subrangle#1{\stackengine{5pt}{}{$\!\scriptstyle #1$}{U}{l}{F}{F}{L}}
\def\sublangle#1{\stackengine{5pt}{}{$\!\scriptstyle #1$}{U}{l}{F}{F}{L}}
\def\subranglestar#1{\stackengine{5pt}{\!\large$^{^*}$}{$\!\scriptstyle #1$}{U}{l}{F}{F}{L}}

\usepackage{multirow}
\newcounter{eqn}

\makeatletter
\newcommand{\putindeepbox}[2][0.7\baselineskip]{{%
    \setbox0=\hbox{#2}%
    \setbox0=\vbox{\noindent\hsize=\wd0\unhbox0}
    \@tempdima=\dp0
    \advance\@tempdima by \ht0
    \advance\@tempdima by -#1\relax
    \dp0=\@tempdima
    \ht0=#1\relax
    \box0
}}
\makeatother

\begin{document}

\title{Laser cooling and magneto-optical trapping of molecules analyzed using optical Bloch equations and the Fokker-Planck-Kramers equation}

\author{J. A. Devlin}\email{j.devlin11@imperial.ac.uk}
\affiliation{Centre for Cold Matter, Blackett Laboratory, Imperial College London, Prince Consort Road, London SW7 2AZ, UK}
\author{M. R. Tarbutt}\email{m.tarbutt@imperial.ac.uk}
\affiliation{Centre for Cold Matter, Blackett Laboratory, Imperial College London, Prince Consort Road, London SW7 2AZ, UK}

\begin{abstract}
We study theoretically the behavior of laser-cooled calcium monofluoride (CaF) molecules in an optical molasses and magneto-optical trap (MOT), and compare our results to recent experiments. We use multi-level optical Bloch equations to estimate the force and the diffusion constant, followed by a Fokker-Planck-Kramers equation to calculate the time-evolution of the velocity distribution. The calculations are done in three-dimensions, and we include all the relevant energy levels of the molecule and all the relevant frequency components of the light. Similar to simpler model systems, the velocity-dependent force curve exhibits Doppler and polarization-gradient forces of opposite signs. We show that the temperature of the MOT is governed mainly by the balance of these two forces. Our calculated MOT temperatures and photon scattering rates are in broad agreement with those measured experimentally over a wide range of parameters. In a blue-detuned molasses, the temperature is determined by the balance of polarization gradient cooling, and heating due to momentum diffusion, with no significant contribution from Doppler heating. In the molasses, we calculate a damping rate similar to the measured one, and steady-state temperatures that have the same dependence on laser intensity and applied magnetic field as measured experimentally, but are consistently a few times smaller than measured. We attribute the higher temperatures in the experiments to fluctuations of the dipole force which are not captured by our model. We show that the photon scattering rate is strongly influenced by the presence of dark states in the system, but that the scattering rate does not go to zero even for stationary molecules because of the transient nature of the dark states.
\end{abstract}

\maketitle

\section{Introduction}

Ultra-cold molecules can be used to test fundamental symmetries, investigate the behavior of strongly-interacting quantum systems, process quantum information, and study collisions and chemistry at low temperature. Direct laser cooling is one way to produce molecules at microkelvin temperatures. Laser cooling has been used to cool several diatomic molecules~\cite{Shuman2010,Hummon2013,Zhelyazkova2014,Lim2018}, and even a polyatomic molecule~\cite{Kozyryev2017}, and magneto-optical traps (MOTs) of SrF~\cite{Barry2014, McCarron2015, Norrgard2016, Steinecker2016}, CaF~\cite{Truppe2017, Williams2017, Anderegg2017} and YO~\cite{Collopy2018} have been demonstrated. Some of the properties of these laser cooled and magneto-optically trapped molecules are similar to their atomic counterparts, while other properties are strikingly different and not fully understood. For example, the spring constants and damping coefficients are both roughly 100 times smaller in molecular MOTs compared to typical alkali atomic MOTs, and the temperature of the molecular MOTs are up to 100 times higher than the Doppler limit, whereas atomic MOTs normally have temperatures close to the Doppler limit.

The main differences between the properties of laser cooled molecules and most laser cooled atoms stems from the different way in which a closed optical cycling transition is achieved. For atoms, it is usual to drive a transition from a ground state of angular momentum $F$ to an excited state of angular momentum $F'=F+1$, referred to as a type-I transition. In this case, there are no dark states, so the atom can scatter the laser light indefinitely. The transition used for laser cooling of molecules is between the lowest rotational level of an electronically excited state, and the first rotationally-excited level of the ground electronic state. This choice avoids decays to multiple rotational states~\cite{Stuhl2008}, but introduces type-II transitions that have $F'\leq F$. In this case, there are dark states amongst the ground state sub-levels. These dark states have a great effect on the scattering rate and associated position and velocity-dependent forces, which in turn influence the properties of a molasses or MOT. 
 
In previous work \cite{Devlin2016}, we studied these position and velocity-dependent forces in 3D molasses and MOTs operating on type-I and type-II transitions, for idealized systems with a single hyperfine ground state and a single hyperfine excited state. In this paper, we present a general method for modeling laser cooling and magneto-optical trapping of molecules, calculate the forces on a real molecule, work out the effect of these forces on the properties of a molasses and a MOT, and compare our findings to experimental results. Some findings are also compared to previous work that used a rate-model approach~\cite{Tarbutt2015,Tarbutt2015b}. We begin by presenting a general model which can be applied to any laser-cooled atom or molecule. Then, we focus our attention on CaF molecules cooled using the {A~$^2\Pi_{1/2}$~--~X~$^2\Sigma^+$~P(1)} transition, since MOTs and molasses of these molecules have recently been studied in depth~~\cite{Truppe2017b, Williams2017, Anderegg2017}.  We examine the predictions of the model and compare them to the results of experiments, first for a molasses, and then for the computationally more complex case of the MOT.

\section{Method}
\label{OBEs}
\subsection{Generalised optical Bloch equations}

We start with a set of optical Bloch equations (OBEs) which describe the time evolution of the internal state of the molecule in the presence of a set of near-resonant laser fields with angular frequencies $\omega_{k}$, and a static magnetic field. To derive the OBEs, we use an identical method to that described in our earlier work \cite{Devlin2016}, generalized to the case of multiple lower and upper hyperfine components. We do not re-derive these equations here, but we do present them in a general form useful to any laser cooling experiment where multiple transitions participate.

The basis states of the system are labeled $\ket{e/g,F_a,M_a}_{\rm s}$ where $e$ denotes an electronically excited state and $g$ a ground electronic state, $F_a$  is the angular momentum, $M_{a}$ is the $z$ projection of angular momentum and $a$ is an index labeling the state. Where there is no ambiguity, we use the shorthand $\ket{e/g,F_a,M_a}_{\rm s}\equiv \ket{e/g,a}_{\rm s}$.  The state $\ket{e/g,a}_{\rm s}$ has energy $\hbar\omega_{e/g,a}$, and it is convenient to define the time-dependent Heisenberg picture operators $\ket{e/g,a}\bra{e/g,b}=e^{-i\omega_{e/g,a}t}\ket{e/g,a}\mathbin{\subrangle{{\rm s}}}e^{i\omega_{e/g,b}t}\mathbin{\sublangle{{\rm s}}\!\!}\bra{e/g,b}$. We find that the expectation values $\langle |e/g,a\rangle  \langle e/g,b|\rangle$ 
evolve according to the following generalized optical Bloch equations:

\begin{align}
\frac{\text{d} \big\langle |g,a\rangle  \langle e,b|\big\rangle }{\text{d} \tau }=& \sum_{\substack{k,q,c}}\frac{i G_k f_{k,q}}{2 \sqrt{2}}e^{i(\overline{\omega}_{e,b}-\overline{\omega}_{g,c}-\overline{\omega}_k)\tau}\mathbin{\sublangle{{\rm s}}\!\!\left\langle g,c\left|\bar{d}_q\right|e,b\right\rangle\subranglestar{{\rm s}}}\langle |g,a\rangle  \langle g,c|\rangle \nonumber\\ 
&-\sum_{\substack{k,q,c'}}\frac{i G_k f_{k,q}}{2 \sqrt{2}}e^{i (\overline{\omega}_{e,c'}-\overline{\omega}_{g,a}-\overline{\omega}_k)\tau} \mathbin{\sublangle{{\rm s}}\!\!\left\langle g,a\left|\bar{d}_q\right|e,c'\right\rangle\subranglestar{{\rm s}}}\langle |e,c'\rangle \langle e,b| \rangle\nonumber\\
   &+\sum_{\substack{q,n}}i (-1)^q \beta _q \,\mathbin{\sublangle{{\rm s}}\!\!\left\langle e,F_b,M_b\left|\bar{\mu}_{-q}\right|e,F_b,n\right\rangle\subrangle{{\rm s}}}\langle |g,F_a,M_a\rangle  \langle e,F_b,n|\rangle \nonumber\\
   &-\sum_{\substack{q,m}}i (-1)^q \beta _q \,\mathbin{\sublangle{{\rm s}}\!\!\left\langle g,F_a,m\left|\bar{\mu }_{-q}\right|g,F_a,M_a\right\rangle\subrangle{{\rm s}}} \langle
   |g,F_a,m\rangle  \langle e,F_b,M_b|\rangle \nonumber\\
   &-\frac{1}{2} \langle |g,a\rangle  \langle e,b|\rangle\,
\label{eq:eg}
\end{align}

\begin{align}
\frac{\text{d} \big\langle |e,a\rangle  \langle e,b|\big\rangle }{\text{d} \tau }=& \sum_{k,q,c}\frac{i G_k}{2 \sqrt{2}}\bigg(f_{k,q} e^{i (\overline{\omega} _{e,b}-\overline{\omega} _{g,c}-\overline{\omega} _k)\tau}  \mathbin{\sublangle{{\rm s}}\!\!\left\langle g,c\left|\bar{d}_q\right|e,b\right\rangle\subranglestar{{\rm s}}} \langle |e,a\rangle  \langle g,c|\rangle \nonumber\\
&-\left(f_{k,q}\right)^* e^{-i (\overline{\omega}_{e,a}-\overline{\omega} _{g,c}-\overline{\omega} _k)\tau}  \mathbin{\sublangle{{\rm s}}\!\! \left\langle g,c\left|\bar{d}_q\right|e,a\right\rangle\subrangle{{\rm s}}} \langle |g,c\rangle \langle e,b| \rangle\bigg)\nonumber\\
&+\sum_{q,n} i (-1)^q \beta _q \, \mathbin{\sublangle{{\rm s}}\!\! \left\langle e,F_b,M_{b}\left|\bar{\mu }_{-q}\right|e,F_{b},n\right\rangle\subrangle{{\rm s}}} \langle |e,F_a,M_a\rangle  \langle e,F_b,n|\rangle \nonumber\\
&-\sum_{q,m} i (-1)^q \beta _q \,\mathbin{\sublangle{{\rm s}}\!\! \left\langle e,F_{a},m\left|\bar{\mu }_{-q}\right|e,F_a,M_{a}\right\rangle\subrangle{{\rm s}}} \langle |e,F_a,m\rangle  \langle e,F_b,M_b|\rangle \nonumber\\
&-\langle |e,a\rangle  \langle e,b|\rangle
\,,
\label{eq:ee}
\end{align}

\begin{align}
\frac{\text{d} \big\langle |g,a\rangle  \langle g,b|\big\rangle }{\text{d} \tau }=&\sum_{k,q,c'}\frac{-i G_k}{2 \sqrt{2}}\bigg(f_{k,q} e^{i (\overline{\omega} _{e,c'}-\overline{\omega}_{g,a}-\overline{\omega} _k)\tau}  \mathbin{\subrangle{{\rm s}}\!\! \left\langle g,a\left|\bar{d}_q\right|e,c'\right\rangle\subranglestar{{\rm s}}} \langle |e,c'\rangle\langle g,b|  \rangle\nonumber\\
&-\left(f_{k,q}\right)^* e^{-i (\overline{\omega} _{e,c'}-\overline{\omega} _{g,b}-\overline{\omega} _k)\tau}   \mathbin{\sublangle{{\rm s}}\!\!\left\langle g,b\left|\bar{d}_q\right|e,c'\right\rangle\subrangle{{\rm s}}} \langle |g,a\rangle  \langle e,c'|\rangle \bigg)\nonumber\\
&+ \sum_{q,n}i (-1)^q \beta _q  \,\mathbin{\sublangle{{\rm s}}\!\!\left\langle g,F_b,M_{b}\left|\overline{\mu} _{-q}\right|g,F_b,n\right\rangle\subrangle{{\rm s}}} \langle |g,F_a,M_a\rangle  \langle g,F_b,n|\rangle\nonumber\\
&-\sum_{q,m} i (-1)^q \beta _q  \,\mathbin{\sublangle{{\rm s}}\!\!\left\langle g,F_a,m\left|\overline{\mu} _{-q}\right|g,F_a,M_{a}\right\rangle\subrangle{{\rm s}}} \langle |g,F_a,m\rangle  \langle g,F_b,M_b|\rangle \nonumber\\
&+\sum_{q,c',c''} \mathbin{\sublangle{{\rm s}}\!\!\left\langle g,a \left|\bar{d}_q\right|e,c'\right\rangle\subranglestar{{\rm s}}} \mathbin{\sublangle{{\rm s}}\!\!\left\langle g,b\left|\bar{d}_q\right|e,c''\right\rangle\subrangle{{\rm s}}} e^{i(\overline{\omega} _{e,c'}-\overline{\omega} _{e,c''}+\overline{\omega} _{g,b}-\overline{\omega} _{g,a})\tau}\langle |e,c'\rangle\langle e,c''| \rangle \, .
\label{eq:gg}
\end{align}

In these equations, the first two terms represent interactions with the lasers, the next two terms capture the effect of an applied magnetic field, and the remaining terms are caused by spontaneous emission. The summations are over the laser frequencies $k$, the polarizations $q=-1,0,1$, all ground states $|g,c\rangle$, all excited state  $|e,c'\rangle$ or $|e,c''\rangle$ and all sublevels $-F_a\leq m\leq F_a$ and $-F_b\leq n\leq F_b$. Spontaneous emission has been introduced via the radiation reaction approximation \cite{Ungar1989}, in which the total electric field interacting with the molecular dipole $\boldsymbol{\hat{d}}$ is written as the sum of the applied electric fields from the lasers and a reaction field $\boldsymbol{\hat{E}}_\textrm{RR}=\tfrac{1}{6\pi\varepsilon_0c^3}\tfrac{\textrm{d}^3}{\textrm{d}t^3}\boldsymbol{\hat{d}}\approx\tfrac{i k^3}{6\pi\varepsilon_0}\boldsymbol{\hat{d}}$, where in the last step we have assumed that the laser frequencies are so close that $\omega_k$ can be replaced by a single $\omega$.  We have made the rotating wave approximation and have assumed that the magnetic field is small enough that all Zeeman shifts are linear. For the specific case of CaF which we consider later, this is a good approximation for fields below 10~G, which is the relevant range for both the molasses and the MOT. To put the equations into natural units, we have used a dimensionless time $\tau = \Gamma t$, and dimensionless angular frequencies $\overline{\omega}_{i} = \omega_{i}/\Gamma$, where $\Gamma$ is the natural decay rate of the excited state. The classical electric field with frequency component $\omega_{k}$ is written as $\boldsymbol{E}_k(\boldsymbol{x},t) = {\cal E}_k \sum_q f_{k,q}(\boldsymbol{x})\epsilon_q^* \cos(\omega_k t)$ and has intensity $I_{k}=\frac{1}{2}c\epsilon_0{\cal E}_k^2$. Here, $\boldsymbol{x}(t)$ is the position of the molecule at time $t$ and the $\epsilon_q$ are the usual spherical basis vectors. The dimensionless parameter $G_{k}$ is defined by $G_{k}=\sqrt{I_k/I_\textrm{sat}}$, where $I_\textrm{sat}=\pi h c \Gamma/(3\lambda^3)$ is the usual expression for the saturation intensity. The applied magnetic field is written as $\boldsymbol{B} = \frac{\hbar\Gamma}{\mu_\textrm{B}} \sum_q \beta_q \epsilon_q^*$ which defines the dimensionless quantity $\beta$. The dimensionless matrix elements of the electric dipole moment operator are
\begin{align*}
\mathbin{\sublangle{{\rm s}}\!\! \left\langle g,a\left|\bar{d}_q\right|e,b\right\rangle\subrangle{{\rm s}}}=&\frac{(-1)^{F_a-M_{a}}\sqrt{2F_b+1}\,\,\mathbin{\sublangle{{\rm s}}\langle g,F_a\|d\|e,F_b\rangle\subrangle{{\rm s}}}}{\sqrt{\sum_c\left|\,\,\mathbin{\sublangle{{\rm s}}\langle e,F_b\|d\|g,F_c\rangle\subrangle{{\rm s}}}\right|^2}}\begin{pmatrix}
 F_a & 1 & F_b \\
    -M_{a} & q & M_{b}
\end{pmatrix}\,, 
\end{align*}
where the sum over $c$ includes all ground states. 
The dimensionless matrix elements of the magnetic moment operator are 
\begin{align*}
\mathbin{\sublangle{{\rm s}}\!\! \left\langle e/g,F_a,m\left|\bar{\mu}_q\right|e/g,F_a,n\right\rangle\subrangle{{\rm s}}}=&-g_{F_a}(-1)^{F_a-m}\sqrt{F_a(F_a+1)(2F_a+1)}
\begin{pmatrix}
 F_a & 1 & F_a\\
    -m & q & n
\end{pmatrix}\, , 
\end{align*}
where $g_{F_a}$ is the magnetic $g$-factor of the state $|e/g,a\rangle$.

We use these equations to calculate how the internal state of a moving molecule evolves over time. The equations may contain far-off-resonant couplings, particularly if there are large energy splittings amongst the ground or excited states. These terms have almost no effect on the results, but can greatly slow down the simulations, so we remove terms oscillating at frequencies greater than a certain threshold, typically $10\Gamma$. In general, for a real multi-level molecular system in which each transition can be excited by several laser frequencies, the molecular operators do not come to a steady state, but continue to vary as a function of time even if the molecule is stationary. If the Hamiltonian is periodic in time, then once any transients relating to the initial conditions die away, the expectation values of the molecular operators also become periodic in time, with the same periodicity as the Hamiltonian \cite{Yudin2016}. In this work, we use rounded values for all frequencies and speeds which sets the periodicity of the Hamiltonian, and hence ensures the molecular operators will eventually reach a periodic quasi-steady state. 

Starting with the population evenly distributed over all ground states, we solve the initial value problem for the ordinary differential equations (\ref{eq:eg}-\ref{eq:gg}) using an explicit Runge-Kutta numerical method, implemented using the \emph{Mathematica} software package, propagating the molecular operators for a large number of time steps ($\tau \approx 1000$). We check how closely the system has reached the quasi-steady state after this initial propagation period by comparing the values of the internal state variables and the velocity-dependent force (defined below) averaged over the next two time periods of the Hamiltonian. A large difference in the calculated values indicates that a longer time period should be used. 
An alternative method of finding the periodic steady state is to calculate the eigenvectors of the propagation matrix \cite{Yudin2016}. This may be faster, though so far in our investigations both solution methods take a similar amount of time.

The quasi-steady state expectation values of the molecular operators are now used to calculate several relevant properties. One is the expectation value of the force operator, $\boldsymbol{\hat{f}}=\textrm{d}\boldsymbol{\hat{P}}/\textrm{d}t=-\nabla\hat{H}$, which may be written as 
\begin{equation}
\boldsymbol{F}\! =\!\langle \boldsymbol{\hat{f}}\rangle\! =\!\!\sum_{k,q,c,c'}\!\!\frac{-h\Gamma G_k}{2\sqrt{2} \lambda}e^{-i (\overline{\omega} _{k}+\overline{\omega} _{g,c}-\overline{\omega}_{e,c'})\tau} \mathbin{\sublangle{{\rm s}}\!\!\left\langle e,c'\left|\bar{d}_q\right|g,c\right\rangle\subrangle{{\rm s}}}\big\langle |e,c'\rangle  \langle g,c|\big\rangle\frac{\partial f_{k,q}(\boldsymbol{x})}{\partial\boldsymbol{x}}+\textrm{c.c.}
\end{equation}
We calculate the force, averaged over one period of the Hamiltonian's oscillation, as the molecule is dragged at constant velocity through the light field. Repeating this for various velocities gives the velocity-dependent force curve, and the gradient of this curve around the equilibrium velocity gives the damping coefficient. We note that this method of determining the velocity-dependent force has been used in several studies on laser cooling of atoms, e.g. \cite{Ungar1989,Molmer1991} and is known to give good agreement with the force obtained from Monte Carlo simulations when the speed is high enough~\cite{Ungar1989}. That comparison suggests that, for the molecular system considered in this paper, the method will be accurate  for all speeds above 0.1~m/s.

Another important quantity is the momentum diffusion tensor, whose components are
\begin{align}
D_{ij}=\frac{1}{2}&\frac{\textrm{d}}{\textrm{d}t}\left(\langle \hat{P}_i\hat{P}_j\rangle-\langle \hat{P}_i\rangle\langle \hat{P}_j\rangle\right)=\textrm{Re}\int_{-\infty}^t\left[\langle \hat{f}_i(\tau)\hat{f}_j(t)\rangle-\langle \hat{f}_i(\tau)\rangle \langle \hat{f}_j(t)\rangle\right]\textrm{d}\tau.\label{eq:fullDiff}
\end{align}
Here $\hat{P}_i$ and $\hat{f}_i$ are the Cartesian components of the momentum and force operator respectively. The momentum diffusion tensor is diagonal if we choose the quantization axis as one of the coordinate axes, and if the laser beams also propagate along a coordinate axis \cite{Ungar1989}. If, furthermore, the laser configuration is identical along all three coordinate axes, then the three diagonal coefficients are all equal, $D_{xx} = D_{yy} = D_{zz} = D$. The diffusion constant includes the effects of (i) the random momentum kicks due to spontaneous emission, (ii) the random momentum kicks due to fluctuations in the difference between the number of photons absorbed from each of the laser beams, and (iii) the fluctuations in the dipole force that arise as the molecule hops between its quantum states in the presence of the intensity and polarization gradients of the light~\cite{Gordon1980,Dalibard1989}. Evaluating Eq.~(\ref{eq:fullDiff}) for a molecule in motion through a complicated light field is a substantial challenge, since the term $\langle \hat{f}_i(\tau)\hat{f}_j(t)\rangle$ involves the expectation value of products of molecular operators at different times of the form $\langle \left(|g\rangle  \langle e|(\tau)\right)\left(|e\rangle \langle g|(t)\right)\rangle$, which are distinct from the expectation values of the operators yielded by solving the OBEs. M\o lmer \cite{Mo/lmer1991} provides a method for calculating $D$ for a stationary atom with a single lower and upper hyperfine state interacting with a single laser beam. However, this method is not straightforward to apply to our case since, in the presence of several laser beams of different frequencies, there is no longer a stationary solution to the OBEs. A later paper \cite{Agarwal1993} provides a method for calculating the first order velocity dependence of the diffusion tensor for a simpler system, which could potentially be adapted to our case. Instead, we choose to approximate the diffusion tensor by a simple expression that ignores stimulated emission and only considers the random nature of spontaneous emission and the corresponding absorption events. In this approximation, the diffusion constant $D_{\rm s}$ is related to the total excited state population $N_e=\sum_{c'}\big\langle |e,c'\rangle  \langle e,c'|\big\rangle$ according to
\begin{align}
D_{\rm{s}}=\frac{1}{3}\hbar^2k^2\Gamma N_e.
\label{eq:diff}
\end{align}
This approximation neglects the extra diffusion caused by the light standing wave pattern, so can only provide a lower limit to the diffusion constant. As with the force, we average the diffusion constant over a period of the Hamiltonian in subsequent calculations of the temperatures and damping constants.

From the force and diffusion constant, we can predict how the fraction of molecules in the phase space element $\textrm{d}\boldsymbol{x}\, \textrm{d}\boldsymbol{v}\,\textrm{d}t$, denoted by  $W(\boldsymbol{x},\boldsymbol{v},t)\textrm{d}\boldsymbol{x}\, \textrm{d}\boldsymbol{v}\, \textrm{d}t$, evolves over time. This evolution is governed by a Fokker-Planck-Kramers equation \cite{Marksteiner1996,Molmer1994}, which for a molecule of mass $m$, moving in 3D is 

\begin{equation}
\frac{\partial W(\boldsymbol{x},\boldsymbol{v},t)}{\partial t}+\sum_{i}v_i\frac{\partial W(\boldsymbol{x},\boldsymbol{v},t)}{\partial x_i}=\sum_{i}\frac{\partial }{\partial v_i}\left(\frac{-F_i(\boldsymbol{x},\boldsymbol{v})}{m}W(\boldsymbol{x},\boldsymbol{v},t)+\frac{D_{ii}(\boldsymbol{x},\boldsymbol{v})}{m^2}\frac{\partial W(\boldsymbol{x},\boldsymbol{v},t)}{\partial v_i}\right).
\label{eq:fpe}
\end{equation}

Using the methods discussed above, we find $F(\boldsymbol{x},\boldsymbol{v})$ and $D_{s}(\boldsymbol{x},\boldsymbol{v})$.
To calculate their effective values for a millimeter-sized cloud of moving molecules, we average them over a cube of size $\lambda$, and also average over all directions of motion. These average values are labeled  $\tilde{F}_i(\boldsymbol{x},v)$ and $\tilde{D}_{s}(\boldsymbol{x},v)$, and they are the ones we use in Eq.~(\ref{eq:fpe}). Due to the symmetric arrangement of the laser beams, the force transverse to the direction of motion averages to zero, leaving only a force in the direction of motion. Thus, the average force can be written as $\tilde{F}(\boldsymbol{x},v)\hat{\boldsymbol{v}}$, where $v$ is the speed and $\hat{\boldsymbol{v}}$ is a unit vector in the direction of $\boldsymbol{v}$.  In the special case where the intensity and magnetic field are uniform over the size scale of the molecular distribution, the position dependence vanishes from Eq.~(\ref{eq:fpe}), which we find reduces to the following equation for the probability density, $W(v,t)$:
\begin{equation}
\frac{\partial}{\partial t}v^2W(v,t)=\frac{\partial }{\partial v}\left(\frac{-\tilde{F}(v)}{m}v^2W(v,t)+\frac{v^2\tilde{D}_{s}(v)}{m^2}\frac{\partial W(v,t)}{\partial v}\right).
\label{eq:fpe3}
\end{equation}

We are often interested in the steady-state solution of this equation, which is
\begin{align}
W(v) & = W_0 \exp\left[m \int_0^{v}\frac{\tilde{F}(u)}{\tilde{D}_s(u)}\textrm{d}u\right].
\label{eq:temp}
\end{align}
The fraction of molecules with speeds between $v$ and $v+dv$ is $W(v)\,4\pi v^2 dv$, and $W_0$ is chosen so that ${\int W(v)\, 4\pi v^2 dv=1}$. If the force is a linear drag force $\tilde{F}(v)=-\alpha v$, and the diffusion constant does not depend on speed, then this integral leads to a Maxwell-Boltzmann speed distribution with temperature $T=\tilde{D}_s/k_B\alpha$. More generally, for any other $W(v)$, we calculate $\overline{v^2}$, the variance in the speed, and so get the equivalent temperature of a Maxwell-Boltzmann speed distribution with that variance,
\begin{align}
T=\frac{m}{3k_B}\int_0^{u}v^2W(v)\,4\pi v^{2}\,\textrm{d}v \, ,
\label{eq:temperatureDefinition}
\end{align}
where $u$ is the upper speed to which the functions $\tilde{F}$ and $\tilde{D}_s$ have been found, chosen to be sufficiently high so as not to affect the distribution $W$.

We can now apply these general equations to the laser cooling of calcium monofluoride. In order to do this, it is helpful to summarize some salient experimental details so we can assess what features of the problem need to be considered. 

\section{Application to laser cooling and magneto-optical trapping of CaF}

\begin{figure}[tb]
\centering
\includegraphics[scale=1]{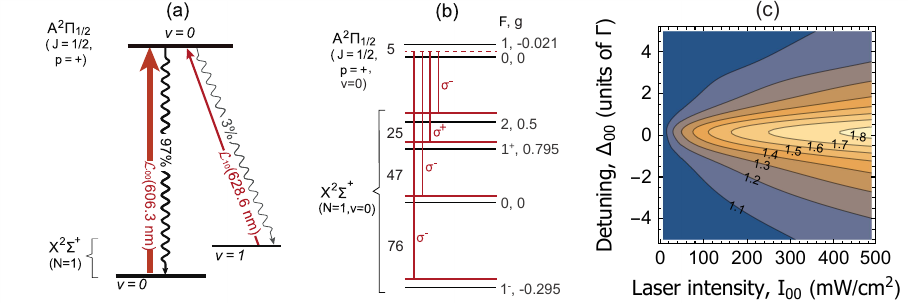}
\caption{(a) Main levels used in the laser cooling of CaF. (b) The hyperfine levels in the $A^2\Pi_{1/2}(v=0)$ and $X^2\Sigma^+(v=0)$ states, and the frequencies of the main cooling laser used to address them. Also indicated is the polarization of each frequency component, as used in the experiments described in \cite{Truppe2017,Williams2017} and in our simulations.  Here, $\sigma^{\pm}$ means that the restoring beams of the MOT drive $\Delta M=\pm 1$ transitions in a coordinate system whose $z$-axis is along the magnetic field. The numbers to the left of the lines are the hyperfine intervals in MHz. (c) The $\eta$ parameter as a function of the intensity and detuning of the cooling laser. $\eta$ is the the total population in $X^2\Sigma^+(v=0)$ and $A^2\Pi_{1/2}(v=0)$ in the case where decays to $X^2\Sigma^+(v=1)$ are forbidden, divided by the total population when decays to this level are allowed.}
\label{fig:CaFenergylevel}
\end{figure}

Our aim is to develop a comprehensive understanding of laser cooling and magneto-optical trapping of molecules. We focus here on the experimental results obtained for dc MOTs and optical molasses of CaF~\cite{Truppe2017, Williams2017, Anderegg2017}, though we expect our methods, and many of our conclusions, to be equally applicable to other molecules and to radio-frequency MOTs~\cite{Norrgard2016}. Figure \ref{fig:CaFenergylevel}(a) shows the relevant energy levels and optical transitions used in the experiments. We use the notation $I_{ij}$ to refer to the total intensity of the lasers addressing the $v=i\rightarrow v'=j$ transition. The main cooling laser drives the  transition $\textrm{A} ^2\Pi_{1/2}(v=0, J=1/2)\leftarrow\textrm{X} ^2\Sigma^{+}(v=0,N=1)$ which has a natural linewidth of $\Gamma=2\pi\times 8.3$ MHz, and a saturation intensity of $I_\textrm{sat}=4.9$~mW/cm$^2$. Population that leaks into $\textrm{X} ^2\Sigma^{+}(v=1,N=1)$ is returned to the cooling cycle using a second laser which we refer to as the repump laser. Additional lasers (not shown in the figure) are used to recover population that leaks to higher-lying vibrational states, but they play such a minor role that we can safely neglect them in the simulations. The cooling laser has four main frequency components to address the four hyperfine ground states shown in Fig.~\ref{fig:CaFenergylevel}(b). The frequency components are derived from a single laser beam; one portion is passed through an electro-optic modulator (EOM) at 74.5 MHz, making three sidebands which address the $F=1^-$, $F=0$ and $F=2$ states, while another portion is passed through an acousto-optic modulator at 48 MHz to address the $F=1^+$ state \cite{Williams2017}. These components are overlapped and directed onto the molecules in three orthogonal pairs of counter-propagating circularly polarized beams in the standard MOT configuration. In the simulations, their intensities are all equal, and their frequencies are $\omega_0+\Delta_{00}+\omega_k$, where $\Delta_{00}$ is a common detuning of the $v=0 \rightarrow v'=0$ laser, $\omega_0$ is the frequency of the $\textrm{X}^2\Sigma^+(v=0, N=1, F=2) \rightarrow \textrm{A}^2\Pi_{1/2}(v=0,J=1/2, F=1)$ transition, and  {$\omega_k=2\pi\times\{-2.90,24.15,72.29,146.00\}$ MHz}. With this definition of $\Delta_{00}$, the simulations predict that the maximum scattering rate at high intensity occurs when $\Delta_{00}=0$, and that there is Doppler cooling when $\Delta_{00}<0$. In the experiment, the EOM produces additional, unwanted, sidebands, but these are detuned from any transition by more than 70~MHz so we neglect them. The repump laser has the same set of four frequency components, but here the common detuning is fixed at zero. 

The parameters used for modeling the MOT and the molasses are summarized in Tab.~\ref{tab:Parameters}. When modeling the MOT, we use laser beams that have Gaussian intensity distributions with $1/e^2$ radii of 8.1~mm, as in the experiment. The magnetic field is $\mathbf{B}=A(\boldsymbol{x}+\boldsymbol{y}-2\boldsymbol{z})$, with $A=15.3$~G/cm. When modeling the molasses, we assume a homogeneous magnetic field randomly oriented with respect to the coordinate axes, and typically use beams with a uniform intensity equal to the true peak intensity. This is valid because the cloud of molecules in the molasses is much smaller than the beam size. When we calculate the capture velocity of the molasses, we consider the full beam profile.

The simulation process is as follows. We first round all angular frequencies and detunings to $\omega_{{\rm min}}$ and round velocities to $\omega_{{\rm min}} \lambda/(2\pi)$, choosing $\omega_{{\rm min}}=10^{-2}\Gamma$ when examining the behavior at high velocity, and $\omega_{{\rm min}}=10^{-3}\Gamma$ if we want more resolution at low velocities. After this procedure, the dimensionless laser frequencies $\omega_k/\Gamma$ are $\bar{\omega}_k=\{-0.36,2.91,8.71,17.59\}$ or $\bar{\omega}_k=\{-0.354,2.909,8.708,17.588\}$. This makes the equations periodic, with period $\mathcal{T}=2\pi/\omega_{{\rm min}}$. Then, for a particular choice of laser intensity, detuning, and applied magnetic field, we solve the OBEs for a molecule moving at constant velocity, $\boldsymbol{v}$, until the quasi-steady-state is reached. From these results we calculate the force and excited-state population averaged over the period $\mathcal{T}$. We do this for random selections of different initial positions, directions of travel, and laser phases, and then average together the results to obtain the mean force and population at this speed $v=|\boldsymbol{v}|$. By repeating this procedure for a range of $v$, we obtain the velocity-dependence of the force and the excited-state population. We use the bootstrap method \cite{BradEandTibshirani1993} to estimate the uncertainty on the mean curve derived this way. Using rounded equations allows the sub-wavelength position-dependent fluctuations in the force and excited state population to be averaged over completely.

\begin{table}[tb]
\caption{Typical experimental parameters used in the simulations. The columns list the intensity of cooling laser, intensity of repump laser, detuning of cooling laser, $1/e^2$ radius of intensity distribution, radial magnetic field gradient. The intensity is the peak intensity due to all four frequency components and all six  beams.}
\begin{center}
\begin{tabular}{|c|c|c|c|c|c|}
\hline
& $I_{00}$ (mW/cm$^{2}$) & $I_{10}$ (mW/cm$^{2}$) & $\Delta_{00}$ ($\Gamma$) & $w$ (mm) & $A$ (G/cm)\\
\hline
MOT & 2.9--466 & 591 & -0.75 & 8.1 & 15.3 \\

Molasses & 2.9--466 & 591 & 2.50 & 8.1 & 0\\

\hline
\end{tabular}
\label{tab:Parameters}
\end{center}
\end{table}

The system described above consists of 28 molecular states, and thus 405 unique coupled equations. This system is small enough that a quasi-steady state solution to the OBEs, and the associated force and diffusion constant, can be computed in around 140~s on a single processor. However, the process of averaging over different trajectories and laser phases described above typically requires hundreds of individual steady-state solutions of the OBEs, which makes it desirable to speed up the calculation. One way to do that is to neglect decay to $X^2\Sigma^+(v=1)$. This approximation has to be treated with caution, because unlike the other states neglected, there is often significant population in this state. Since the repump light is on resonance, we do not expect the $A^2\Pi_{1/2}(v=0)\leftarrow X^2\Sigma^+(v=1)$ transition to contribute directly to the position-dependent or velocity-dependent forces, but it does contribute indirectly by altering the populations of the various states. In particular, neglecting $X^2\Sigma^+(v=1)$ will lead to an overestimate of the populations in $X^2\Sigma^+(v=0)$ and $A^2\Pi_{1/2}(v=0)$, which in turn leads to an overestimate of the force and the diffusion constant. We have investigated this by solving the rate equations for the system \cite{Tarbutt2015b} with and without the $X^2\Sigma^+(v=1)$ levels. For this investigation, we fix $I_{10}$ and vary $I_{00}$ and $\Delta_{00}$ over wide ranges. In all cases, we find that including the $X^2\Sigma^+(v=1)$ levels reduces the population in $A^2\Pi_{1/2}(v=0)$ and $X^2\Sigma^+(v=0)$ by a common factor $\eta(I_{00},\Delta_{00})$. In light of these observations, we suggest that fairly accurate results can be found by solving the OBEs without including the $X^2\Sigma^+(v=1)$ levels, and then dividing both the force and the excited state population by the correction factor $\eta$. The value of $\eta$ is plotted as a function of $I_{00}$ and $\Delta_{00}$ in Fig.~\ref{fig:CaFenergylevel}(c). To validate this approach, we solve the OBEs for a few specific parameters both with and without the $v=1$ levels. We find that when $\Delta_{00}=2.5\Gamma$ (as used in the molasses), the $\eta$-scaled force and excited state population curves agree very well with the full simulations at both high and low values of $I_{00}$. The agreement is also good when $\Delta_{00}=-0.75\Gamma$ (as used in the MOT) and the intensity is low, but less good at higher intensity. For the highest $I_{00}$ used, the $\eta$-scaled excited-state population is accurate, but the $\eta$-scaled force is 60\% lower than given by the full simulations. This underestimate should be kept in mind when considering simulations of the MOT at the highest intensities.

\section{Forces and excited-state populations in the CaF MOT and molasses}

\begin{figure}[tb]
\begin{center}
{\includegraphics{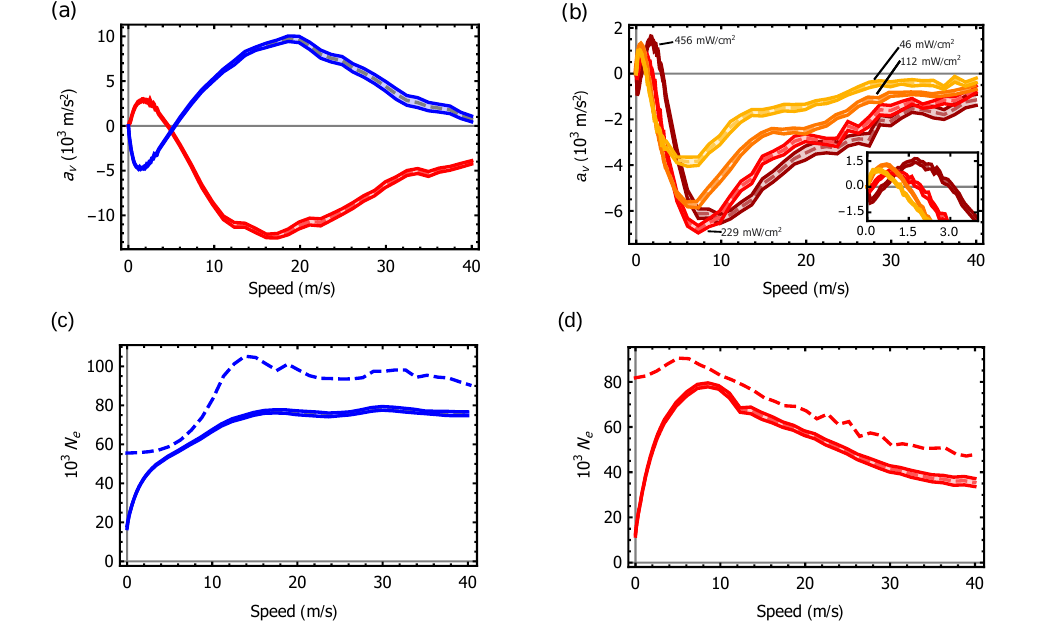}}
\caption{Acceleration and excited-state population as a function of speed. For each curve, the shaded band indicates the 67\% confidence interval determined from the distribution of multiple simulations. (a) Acceleration curves for $I_{00}=456$ mW/cm$^2$. Blue curve (positive acceleration at higher speeds): $\eta=1.29$, $\Delta_{00}=2.61\Gamma$; red curve (negative acceleration at higher speed): $\eta=1.32$, $\Delta_{00}=-2.39\Gamma$. (b) Acceleration curves for $\Delta_{00}=-0.64\Gamma$ and for a series of intensities, $I_{00}$; dark red: 456~mW/cm$^2$, $\eta=1.65$; red: 229~mW/cm$^2$, $\eta=1.48$; orange: 112~mW/cm$^2$, $\eta=1.32$; yellow: 46~mW/cm$^2$, $\eta=1.16$. (c) Steady-state excited-state fraction when $I_{00}=456$ mW/cm$^2$, $\eta=1.45$, and $\Delta_{00}=2.61\Gamma$. (d) Excited-state fraction when $I_{00}=228$ mW/cm$^2$, $\eta=1.48$, $\Delta_{00}=-0.64\Gamma$. In (c,d), the solid curve is calculated from the optical Bloch equations, while the dashed line is the prediction of a rate equation model~\cite{Tarbutt2015}.}
\label{fig:basicPhysics}
\end{center}
\end{figure}

The essential properties of a CaF MOT and molasses can be understood from the velocity-dependence of the acceleration and the excited-state probability, particularly at low velocity. Figure \ref{fig:basicPhysics}(a) shows the acceleration in the direction of the velocity, $a_\textbf{v}(v)$, as a function of the speed of a CaF molecule, for both red-detuned and blue-detuned light. The magnetic field is set to zero, and the other parameters are given in the caption. For high speeds the acceleration is negative for red-detuned light and positive for blue-detuned light, corresponding to normal Doppler cooling or heating. At lower speeds, polarization-gradient forces dominate over Doppler forces and the acceleration changes sign. We note that, despite the complexity of the CaF system, the force curve is very similar to those found for type-II systems with just one ground state and one excited state~\cite{Devlin2016}, and we conclude that the mechanisms at work are the same as for those simpler systems~\cite{Devlin2016, Weidemuller1994, Shahriar1993}. For a stationary molecule excited on a type-II transition between integer valued\footnote{If $F$ is half-integer, the transition from $F$ to $F'=F$ is only dark in circularly polarized light \cite{Berkeland2002}.} hyperfine levels, there is one dark state when $F=F'$, and two dark states when $F=F'+1$. A moving molecule will tend to be pumped into a dark state near the intensity anti-nodes, where the pumping rate is highest, and will tend to make a non-adiabatic transition back to a bright state close to the nodes, where the splitting between bright and dark states, arising from the ac Stark shift, is smallest. For blue-detuned light, the ac Stark shift is positive, so the bright states have higher energy at the anti-nodes than at the nodes. Thus, the molecule will continually lose energy to the light field, leading to a cooling force. For red-detuned light, the sign of the ac Stark shift is reversed, so the molecule continually receives energy from the light field. In Fig.~\ref{fig:basicPhysics}(a), the speed where the force crosses zero is around 5 m/s. This is the rms speed of a 60 mK Maxwell-Boltzmann distribution, giving an approximate temperature scale where the Doppler and polarization-gradient forces are balanced.

Figure \ref{fig:basicPhysics}(b) shows how the acceleration curve changes as the intensity is reduced. Here, the detuning is negative and close to that typically used for the MOT. Both the range and magnitude of the sub-Doppler force is reduced as the intensity is lowered. This explains the experimental observation that lowering the intensity lowers the temperature.
 
Figures \ref{fig:basicPhysics}(c,d) show the excited state fraction as a function of speed for a positive detuning close to the one used in the molasses (part c), and  for a negative detuning close to the one used for the MOT (part d). Here, we compare the population found by solving the OBEs (solid lines) to the predictions of a rate model (dashed lines), described fully in Ref.~\cite{Tarbutt2015}. At all velocities, the OBEs predict a lower excited state fraction, and hence a lower scattering rate, than predicted by the rate model, indicating that transient dark states play an important role in reducing the scattering rate. Close to zero velocity, the excited state fraction dips even further, as the molecule optically pumps into the dark states. The excited state fraction does not quite drop to zero at zero velocity, as would be expected for an isolated type-II transition driven by a single laser frequency. This behavior can be understood with the help of the level structure shown in Fig.~\ref{fig:CaFenergylevel}(b). Population cannot build up in $F=0$ since regardless of the laser polarization this can always be excited to $F'=1$. The presence of the nearly degenerate pair of excited hyperfine states $F'=0$ and $F'=1$ means that any molecules in $F=1^-$ or $F=1^+$ can always interact with the elliptical light field formed by the superposition of three pairs of $\sigma^+\sigma^-$ beams whose phases are not controlled. Population might be expected to build up in $F=2$, since this can only be excited to $F'=1$ and so it appears to be an isolated type-II transition. However when multiple lasers with different frequencies and polarizations drive the same type-II transition, the states that are dark to one laser beam are, in general, bright to the other. Except in certain special cases, for instance if one of the two fields driving the $F\rightarrow F-1$ transition is circularly polarized and the other linear, there is no guaranteed time-independent orientation of the dipole which is simultaneously orthogonal to all of the applied frequency components of light. The molecule therefore never decouples completely from the light field, even at zero velocity.

It is also worth noting that as well as dark states formed between Zeeman sub-levels of a particular hyperfine state, there can also be Raman dark states which are superpositions of two or more Zeeman sub-levels from different hyperfine states. The laser sidebands are phase-coherent, so we can expect these Raman dark states to be stable. To assess their importance, we artificially destabilize these dark states by adding a term to the right hand side of Eq.~\eqref{eq:gg} of the form $-\gamma(1-\delta(\omega_a-\omega_b))\langle |g,a\rangle  \langle g,b|\rangle$, where $\delta(x)$ is the Kronecker delta function and we set $\gamma=10$, so that coherences between hyperfine levels are rapidly damped away. We find that this increases the excited state fraction by about 20\%, reduces the range of the polarization-gradient force by about 50\%, and reduces the magnitude of this force by 33\%, without changing its slope at low velocities. This shows that optical pumping into Raman dark states, and non-adiabatic transitions out of these states, is an important part of the polarization-gradient cooling mechanism in this multi-level system.

\section{Simulations of the molasses}
\subsection{Damping constant and capture velocity}

\begin{figure}[tb]
\begin{center}
\includegraphics{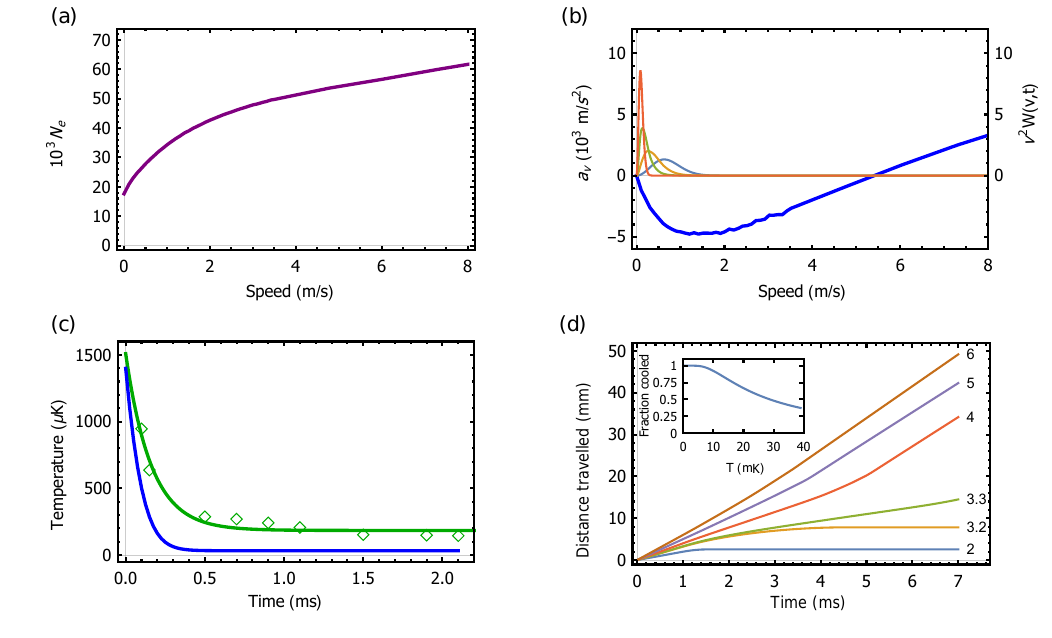}
\caption{Predicting the temperature of a CaF molasses. For all curves, parameters are $I_{00}=456$~mW/cm$^2$, $\eta=1.29$, and $\Delta_{00}=2.61\Gamma$. (a) Excited state population as a function of speed.  (b) Thick, negative blue curve: Acceleration parallel to velocity versus speed. Other curves: speed distribution, $W(v,t)$, for $t=\{0,0.1,0.2,1\}$~ms. The initial distribution has a temperature of $1.4$~mK, and the distributions get narrower with time. (c) Lower, blue curve: simulated temperature as a function of molasses cooling time. Green diamonds: experimental data. Upper green line: fit of experimental data to $T=T_{{\rm final}} + (T_{{\rm init}}-T_{{\rm final}})e^{-t/\tau_{{\rm cool}}}$.  (d) Displacement versus time for molecules of various initial speeds, showing which are effectively cooled to rest. Curves are labeled by the initial speed. Inset: fraction of molecules captured in the molasses as a function of starting temperature.}
\label{fig:molassescooling}
\end{center}
\end{figure}

We now compare the results of our simulations to experimental data. We start by considering cooling in the molasses, since this is easier to simulate than the MOT. We simulate the case where $\Delta_{00}=2.61\Gamma$ and $I_{00}=456$~mW/cm$^2$, and assume that the molasses is loaded from a thermal distribution with a temperature of 1.4~mK, which is common in the experiments. Figures \ref{fig:molassescooling}(a,b) show the excited-state population ($N_e$) and the acceleration ($a_v$), as functions of the speed. For a linear damping, $a_v=-\alpha v/m$, we would expect the temperature of an initially hot sample to decay exponentially with a $1/e$ time constant of $t_d=m/2\alpha$. Taking the gradient of $a_v$ near zero velocity, we find {$\alpha=10.2(0.5)\times10^3$ s$^{-1}$}, implying a characteristic damping time of {$t_d=49(2)$ $\mu$s}. However, the $a_v$ curve is only linear for speeds below about 0.5~m/s, while the initial velocity distribution extends to significantly higher speeds. To get a better estimate of the damping time we solve Eq.~\eqref{eq:fpe}, taking the diffusion constant given by Eq.~\eqref{eq:diff}, and the $a_v$ and $N_e$ curves shown in Fig.~\ref{fig:molassescooling}(a,b). The resulting velocity distributions are shown for four different times in Fig.~\ref{fig:molassescooling}(b). From distributions such as these, we obtain the predicted temperature as a function of time, which is shown by the blue line in Fig.~\ref{fig:molassescooling}(c). The figure also shows recent experimental data obtained at an intensity of $I_{00}=456$~mW/cm$^2$, and under conditions where the magnetic field is carefully zeroed and the laser detuning is switched rapidly from the MOT to the molasses phase. The predicted damping time is 101(1) $\mu$s, fairly close to the measured time of 160(30) $\mu$s. The predicted final temperature is 30 $\mu$K, about 5 times lower than the measured value of 144 $\mu$K. We attribute this discrepancy to the incomplete treatment of the diffusion constant, which omits the fluctuations in the dipole force. For a type-I transition in a one-dimensional lin$\perp$lin configuration, the diffusion related to the fluctuations of the dipole force is approximately $\Delta^2/\Gamma^2$ times greater than $D_s$ \cite{Dalibard1989}. There are intensity gradients in the 3D molasses, so we can expect the same mechanism to be present. If the type-II transition exhibits similar scaling, this would lead to a seven-fold increase in the temperature, bringing it closer to the experimentally observed result. A more thorough treatment of the diffusion constant for this multilevel system in 3D is desirable. We also note that at 30 $\mu K$, the thermal de Broglie wavelength of 40~nm is approaching the wavelength of the light, so the validity of the classical treatment of the molecular position and momentum begins to break down.  A full quantum treatment of the position and momentum of the molecule is called for to analyze fully these lowest temperature cases.

Next, we consider the capture velocity of the molasses. We calculate $a_v(v)$ for many different intensities, and then use the known laser beam profile to generate the map $a_v(v,r)$, where $r$ is the displacement from the center of the molasses. We then consider a CaF molecule traveling outwards from the center of the molasses, and plot its displacement as a function of time. Figure \ref{fig:molassescooling}(d) shows a series of these curves for several different starting velocities. Here, the parameters are $\Delta_{00}=2.61\Gamma$ and $I_{00}=456$ mW/cm$^2$. The maximum speed at which the molasses is able to bring the molecule to rest is $v_c=3.1$~m/s. To estimate the fraction of molecules that can be cooled, we integrate a Maxwell-Boltzmann speed distribution up to $v_c$ for a range of temperatures. The results are plotted in the inset to Fig.~\ref{fig:molassescooling}(d). For initial temperatures below 5~mK all the molecules are cooled by the molasses.  At higher temperatures, a fraction of them escape from the molasses before they can be cooled.

\subsection{Dependence of temperature on intensity, detuning and magnetic field}

\begin{figure}[tb]
\begin{center}
{\includegraphics{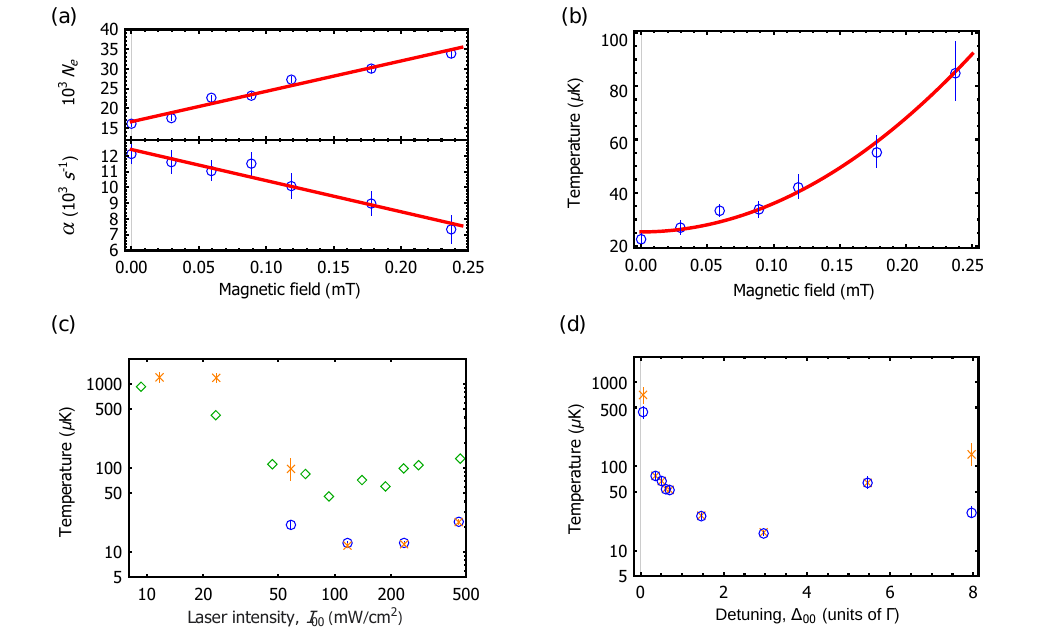}}
\caption{ (a) Excited state population, $N_e$, at zero velocity, and the damping constant, $\alpha$, as a function of magnetic field. (b) Molasses temperature as a function of magnetic field. Parameters are $I_{00}=456$ mW/cm$^2$, $\eta$=1.29 $\Delta_{00}=2.61\Gamma$. Red line: quadratic fit to blue points. (c) Molasses temperature as a function of cooling laser intensity, $I_{00}$, for $\Delta_{00}=2.61\Gamma$. Values of $\eta$ for points running left to right are $\{1.01,1.02,1.05,1.10,1.17,1.29\}$. A thermal distribution with a temperature of 1~mK is loaded into the molasses. Blue circles: steady-state temperatures; Orange crosses: temperatures after 5~ms of cooling; Green diamonds: data from reference \cite{Truppe2017} with $\Delta_{00}=2.5\Gamma$. (d) Molasses temperature as a function of laser detuning, $\Delta_{00}$, with $I_{00}=467$ mW/cm$^2$. Values of $\eta$ for points running left to right are $\{ 1.82,1.80,1.77,1.74,1.71,1.47,1.25,1.12,1.07\}$. A thermal distribution with a temperature of 1.4~mK is loaded into the molasses. Blue circles: steady-state temperatures; Orange crosses: temperatures after 5~ms of molasses cooling.}
\label{fig:molassessensitivity}
\end{center}
\end{figure}

To investigate the effects of a background magnetic field, we apply the same procedure as described above, with a randomly oriented uniform field applied. We first solve the OBEs to calculate $a_v(v)$ and $D_s(v)$ for various magnetic field strengths. Figure \ref{fig:molassessensitivity}(a) shows the linear slope of the acceleration close to zero velocity, $\alpha= -\left(\frac{d a_v}{d v}\right)_{v=0}$, and $N_e(0)=\frac{3D_s(0)}{\hbar^2k^2\Gamma}$ as a function of the absolute value of the magnetic field. We see that the damping decreases linearly with magnetic field strength, whereas the diffusion constant increases linearly over the range considered. If we simply use these linear gradients, along with $k_{\rm B} T =D_{s}/\alpha$, we would expect the temperature $T$ in $\mu K$ as a function of the magnetic field $B$ in mT to be $T(B)=21 + 130B + 200B^2+310B^3$, where terms of order $B^4$ have a negligible contribution over the range of $B$ considered here. To investigate the effects of the full velocity dependence of $a_v(v)$ and $D_s(v)$, we find the steady-state temperature  using Eq.~\eqref{eq:fpe3}. The results are shown in Fig.~\ref{fig:molassessensitivity}(b), and show a very different dependence to the one expected from the linear approximation made above. The temperature fits well to a purely quadratic dependence on $B$, with a curvature of 1070(50) $\mu$K/(mT)$^2$. The experiment also found a quadratic dependence, but with the larger coefficient of 5740(30) $\mu$K/(mT)$^2$. The discrepancy between predicted and measured curvatures can again be explained by a systematic underestimation of the temperature because of the missing part of the diffusion constant, again by a similar factor of 5.6. 

Figure \ref{fig:molassessensitivity}(c) shows how the molasses temperature depends on the laser intensity, $I_{00}$. Here, the blue circles give the predicted steady-state temperature, the orange crosses give the temperature after 5~ms of molasses cooling, and the green diamonds are the experimental data points. At high intensity, we see once again a five fold underestimation of the temperature. As the intensity is decreased, both in the simulation and in the experiment, the temperature is reduced. This is because the low velocity part of $a_v$ is independent of intensity, but lowering the intensity reduces the excited state fraction and hence the diffusion constant. In both the experiment and the simulation, the optimum intensity is around 100~mW/cm$^2$. At lower intensities than this, the temperature rises, and we see a difference between the predicted steady-state temperature and its value after only 5~ms of cooling. This rise in temperature at low intensity occurs because the velocity range of the sub-Doppler force is so low that it can only cool the slowest molecules, so the value of $a_v$ averaged over the velocities of the molecules is reduced. At very low intensities, $I_{00} \lesssim 20$~mW/cm$^2$, we see that the temperature of the molasses after 5 ms is close to 1~mK, which is the temperature at which the molasses was loaded, implying that the molasses no longer cools the distribution at all. In fact, for these intensities the entire molecular distribution is heated to velocities greater than 5 m/s and we could not find a steady-state temperature (which is why there are no blue circles plotted for these low intensities). 

Figure \ref{fig:molassessensitivity}(d), shows how the temperature depends on detuning. The temperature decreases rapidly as the detuning is increased from zero, reaches a minimum around $\Delta_{00}=3\Gamma$, exactly as in the experiments, and then gradually increases with further increases in detuning.  As can be seen from Fig.~\ref{fig:CaFenergylevel}(b), some of the laser frequency components become resonant with some hyperfine levels as the detuning is scanned over this range. Surprisingly, we do not see any structure in the plot of temperature versus detuning that reflects the hyperfine structure of the molecule. Indeed, at the detuning where the temperature is minimized, the  lowest frequency component of the laser is near-resonant with the upper $F=1$ level, but this does not appear to raise the temperature.

\section{Simulations of the MOT}

\subsection{Method for modelling the MOT}

\begin{figure}[tb]
\begin{center}
\begin{tabular}{cc}
   \putindeepbox[1pt]{\includegraphics{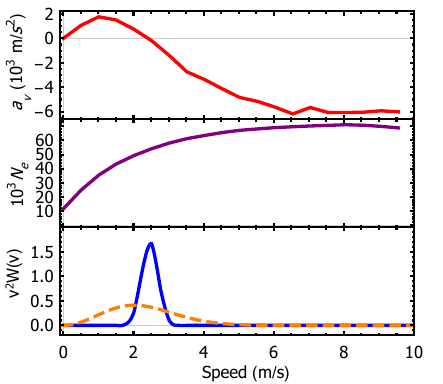}}
\end{tabular}
\caption{Top: $a_v(v)$ for a molecule at the center of the MOT. Middle: $N_e(v)$ for a molecule at the center of the MOT. Bottom: Blue line is $v^2W(v)$ calculated using $a_v(v)$ and $N_e(v)$, orange dashed line is a Maxwell-Boltzmann speed distribution of temperature T=14.5~mK. Parameters are $I_{00}=234$~mW/cm$^2$, $\Delta=-0.64 \Gamma$ and $\eta=1.49.$}
\label{fig:motPropertiesSimple}
\end{center}
\end{figure}

The MOT is more complicated to model than the molasses, because the position-dependence of the magnetic field and laser intensity modify the force and excited state population. This means that both $a_v$ and $D_s$ depend on the axial and radial displacement from the MOT center, as well as the speed. Before discussing how we deal with the additional complication, let us first focus on the behavior of a molecule at the very center of the MOT. The top two panes in Fig.~\ref{fig:motPropertiesSimple} plot $a_v$ and $N_e$ at the center of a MOT with $I_{00}=234$ mW/cm$^2$ and $\Delta_{00} =-0.64 \Gamma$. The lower pane shows the steady-state molecular speed distribution found from these $a_v(v)$ and $N_e(v)$ using Eq.~\eqref{eq:temp}. The change in sign of the damping force at a speed of around 2.5 m/s, where Doppler cooling turns into Sisyphus heating, leads to a peak in the speed distribution at this point. The temperature of this distribution, calculated using Eq.~\eqref{eq:temp} and Eq.~\eqref{eq:temperatureDefinition}, is 14.5~mK, far higher than the Doppler-limited temperature of 570 $\mu$K for this detuning and intensity. The distribution looks nothing like a Maxwell-Boltzmann distribution at this temperature, which is shown, for comparison, by the orange dashed line. Nevertheless, and remarkably, we find that when a collection of molecules with speeds drawn from this distribution expands freely, their rms width $\sigma$ increases as a function of time according to $\sigma^2(t)=\sigma_0^2+k_BTt^2/m$, with the same temperature $T$ as found above.

\begin{figure}[tb]
\begin{center}
\includegraphics{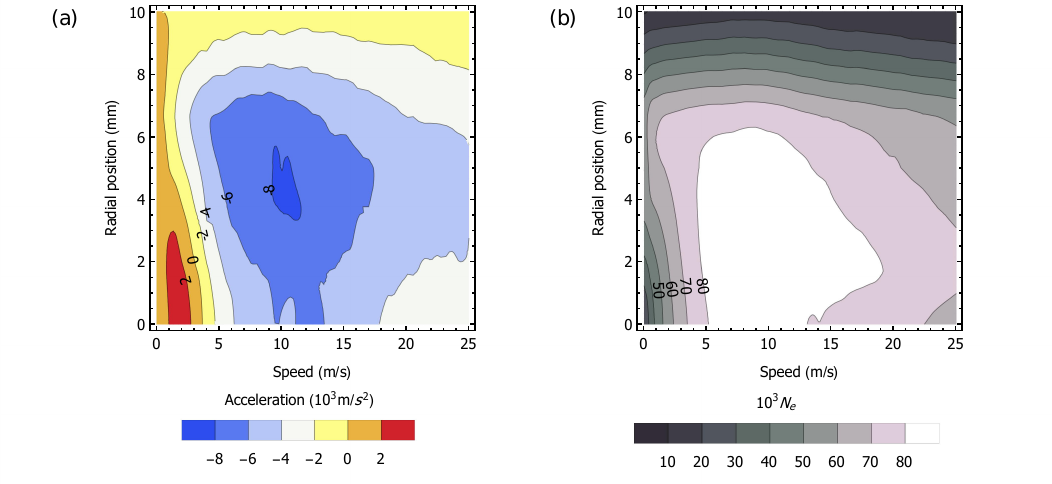}
\caption{(a) $a_v$ and (b) $N_e$ as a function of displacement along the $x$ axis and speed $v$. Parameters are: $I_{00}=468$ mW/cm$^2$, $
\Delta=-0.64\Gamma$. $\eta$ ranges from $1.65$ in the center to $1.09$ at $x=10$~mm.}
\label{fig:radialField}
\end{center}
\end{figure}

Next we consider what happens away from the center of the MOT. Figure \ref{fig:radialField} shows contour plots of $a_v(x,v)$ and $N_e(x,v)$ as a function of speed $v$ and displacement $x$ along the $x-$axis. In (a), cooling is indicated in blue-yellow, and heating in red-orange. In (b), darker regions indicate low excited-state probability, and lighter regions indicate higher probability. Focusing on the heating at low velocity, we see that as the distance from the center of the MOT increases, the gradient of the force curve at zero velocity decreases, the maximum value of the force also decreases, and the range over which the polarization-gradient heating acts is reduced. All are due to the increasing magnetic field which reduces the effectiveness of the polarization-gradient force. Around 5~mm from the center the heating force has almost vanished, and the force curve is dominated by Doppler cooling. At even larger distances, the decreasing laser intensity reduces the Doppler cooling force.  Figure \ref{fig:radialField} (b) shows that as the displacement from the MOT center increases towards $\sim 3$~mm, the excited-state population increases for all speeds, but increases most strongly when the speed is low. Since the temperature in the MOT is primarily determined by the zero crossing point of $a_v (v)$, these changes in $N_e$ do no have much effect on the temperature. They affect the total scattering rate of the MOT, tending to make it brighter than the equivalent molasses. Once the displacement grows larger than around 7.5~mm, the decreasing laser intensity strongly reduces the excited state fraction. Experimentally, it is found that for the choice of parameters used here, the radial rms width of the cloud is about 2.0~mm. Therefore, for accurate predictions of the MOT properties, we must account for the position dependence of $a_v$ and $N_e$.

Using our calculation of $a_v(x,v)$ and $N_e(x,v)$, it is possible to calculate the complete phase-space distribution by numerically solving Eq.~(\ref{eq:fpe}). Because this is difficult, we adopt a simpler method where we assume that the density distribution is already known from experiment, and that momentum and position are uncorrelated. In this case, the phase-space distribution is $W(\boldsymbol{x},\boldsymbol{v},t)=U(\boldsymbol{x})P(\boldsymbol{v},t)$. Integrating Eq.~(\ref{eq:fpe}) over the spatial coordinates, and using the fact that $U(\boldsymbol{x})\rightarrow 0$ as $\{x,y,z\} \rightarrow \pm \infty$, we obtain 
\begin{equation}
\frac{\partial}{\partial t}v^2P(v,t)=\frac{\partial }{\partial v}\left(\frac{-\mathcal{F}(v)}{m}v^2P(v,t)+\frac{v^2\mathcal{D}_{S}(v)}{m^2}\frac{\partial P(v,t)}{\partial v}\right),
\label{eq:fpeMOT}
\end{equation}
where 
\begin{align}
\mathcal{F}(v)&=\int_{-\infty}^{\infty}\int_{-\infty}^{\infty}\int_{-\infty}^{\infty} U(\boldsymbol{x})\tilde{F}(\boldsymbol{x},v)\textrm{d}x \textrm{d}y\textrm{d}z \,,\\
\mathcal{D}_{s}(v)&=\int_{-\infty}^{\infty}\int_{-\infty}^{\infty}\int_{-\infty}^{\infty} U(\boldsymbol{x})\tilde{D}_{s}(\boldsymbol{x},v)\textrm{d}x \textrm{d}y\textrm{d}z\, ,
\end{align}
are the position-weighted averages of the force and diffusion constant. Note that, similar to Eq.~(\ref{eq:diff}), $\mathcal{D}_{s}$ is related to the position-weighted excited-state population, $\mathcal{N}_{e}$, as $\mathcal{D}_{s}(v) = \hbar^2 k^2 \Gamma \mathcal{N}_{e}(v)/3$.

The force is the sum of a trapping term, which is antisymmetric under $\boldsymbol{x}\rightarrow\boldsymbol{-x}$, and a damping term which is symmetric. Assuming $U(\boldsymbol{x})$ is symmetric, we see that the trapping term makes no contribution to the integral and we only need to use the damping term. To perform the integrals, we find $\tilde{F}(\boldsymbol{x},\boldsymbol{v})$ and $\tilde{D}_s(\boldsymbol{x},\boldsymbol{v})$ by repeatedly solving the OBEs using the methods discussed previously, where the intensity and magnetic field take their values at position $\boldsymbol{x}$ in the MOT. We take $U(\boldsymbol{x})$ to be the density distribution measured in the experiment, 
\begin{align}
U(\boldsymbol{x})&=\frac{1}{\sigma_x^2\sigma_z(2\pi)^{3/2}}e^{-\frac{x^2+y^2}{2\sigma_x^2}} e^{-\frac{z^2}{2\sigma_z^2}}\, ,
\end{align}
where $\sigma_x$ and $\sigma_z$ are the measured rms widths in the radial and axial directions. We then calculate the weighting integrals to arrive at $\mathcal{F}$ and $\mathcal{D}_{s}$. As we shall see, this approach reproduces many of the the observed properties of the MOT.

\subsection{MOT properties}

\begin{figure}[htb]
\begin{center}
\includegraphics{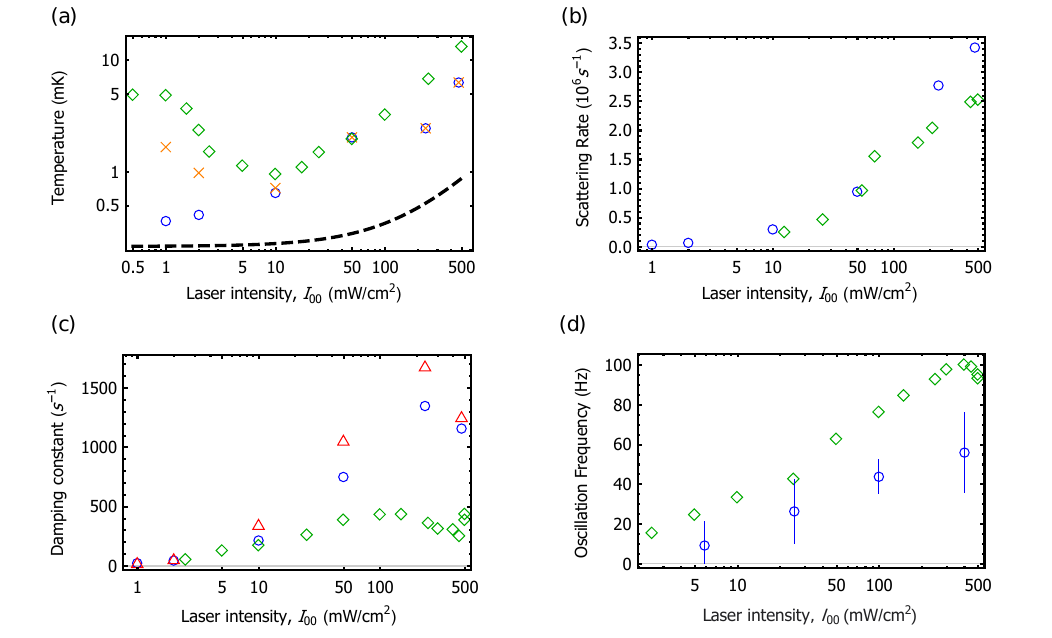}
\caption{A comparison between experimentally measured properties of the CaF MOT \cite{Williams2017} (green diamonds) and the numerical simulations (other points) as a function of $I_{00}$. The detuning is $\Delta=-0.64\Gamma$ in the simulations and $\Delta=-0.75\Gamma$ for the experimental data. (a) Temperature. Blue circles show the predicted steady-state temperature. Orange crosses show the temperature following the intensity ramp performed in the experiment (see main text). The black dashed line is the predicted Doppler cooling limit (see Eq.(14) of Ref.~\cite{Williams2017}). (b) Photon scattering rate. (c) Damping rate for radial oscillations. Red triangles: damping rate inferred from the gradient of the acceleration curve at zero velocity. Blue circles: damping rate inferred from the time constant for the temperature to approach equilibrium. (d) Frequency of radial oscillations.}
\label{fig:motProperties}
\end{center}
\end{figure}

Measurements of the MOT temperature as a function of $I_{00}$~\cite{Williams2017} are plotted as green diamonds in Fig.~\ref{fig:motProperties}(a), while the calculated steady-state temperatures found by solving Eq.~(\ref{eq:fpeMOT}) are shown as blue circles in this figure. We see fairly good agreement for the points between 10 and 500~mW/cm$^2$. Below 10~mW/cm$^2$, the predicted steady-state temperature decreases with decreasing intensity, whereas the measured points trend upwards. Here, at least part of the discrepancy might be caused by the long thermalization time of the MOT at these low intensities. In the experiments, the intensity is first held at 468~mW/cm$^2$ to load the MOT, then ramped down to its new value over 20~ms, then held at this value for 5~ms, before measuring the temperature. We replicate this procedure numerically by first calculating the velocity distribution for a MOT with $I_{00}=468$~mW/cm$^2$, then calculating how this distribution is modified when $I_{00}$ is varied as in the experiment. The results of these simulations are shown as orange crosses in the figure. At high intensities, this procedure gives the same temperature as in the steady-state (the orange crosses lie on top of the blue circles). At lower intensities, we see that the 20~ms cooling ramp and/or 5~ms hold time are not slow enough for the temperature to reach the steady-state. As a result, the temperature rises at low $I_{00}$. The agreement between simulation and experiment is reasonably good over the whole range of intensities (spanning 3 orders of magnitude). Notably, our model gives far better predictions of the temperature than the Doppler-limited temperature predicted by a rate equation model, which is plotted as the black dashed line in the figure.

Once Eq.~(\ref{eq:fpeMOT}) has been solved in the steady state to give $P(v)=\lim_{t\to\infty} P(v,t)$ at each intensity, it can be used to calculate the scattering rate, which is 
\begin{align}
R_{\rm sc} = \Gamma \int_{0}^{\infty} \mathcal{N}_e(v) P(v)\,4\pi v^{2}dv\,.
\end{align}
This scattering rate is plotted in Fig.~\ref{fig:motProperties} (b), and is seen to agree excellently with the measured scattering rate. By comparison, the rate equation model, which cannot capture optical pumping into transient dark states, overestimates the scattering rate by a factor of 2~\cite{Williams2017}. 

Experimentally, it is found that following a radial push, the cloud executes damped harmonic motion, with the position, $r$, of the center of the cloud, following the equation
\begin{align*}
r''=-\omega^2r-b r'\, .
\end{align*} 
Here, $\omega$ is the angular oscillation frequency, and $b$ is the damping rate. The green diamonds in Fig.~\ref{fig:motProperties} (c) shows the measured damping rate as a function of intensity. We can determine the damping rate from the simulations in two different ways. Firstly, we can calculate the slope of $\mathcal{F}(v)/m$ around the zero crossing velocity $v_0$, $\alpha'= -\left(\frac{d a_v}{d v}\right)_{v=v_0}$, which should be equal to $b$. The results obtained this way are plotted as red triangles in Fig.~\ref{fig:motProperties} (c). Alternatively, we can perturb the steady state distribution $P(v)$ by changing it to $P(v-v_{\rm push})$, where $v_{\rm push}=2.3$~m/s is the speed given to the molecules in the measurements of the damped oscillations. We then find $\tau_{\rm d}$, the $1/e$ time constant with which the temperature is damped, which should be related to the velocity damping constant by $b=1/(2\tau_{\rm d})$. These points are plotted as blue circles in this figure. Both methods yield similar results, and both have the same qualitative shape as the experimentally measured distribution. At low intensities the agreement between the simulations and the experiment is good. However, the simulations overestimate the damping constant by a factor of 3--5 at higher intensities. While not perfect, the agreement here is again better than the rate equation model, which overestimates the damping constant by a factor of 2-3 at low intensities and 5-10 at higher intensities. 

Figure \ref{fig:motProperties} (d) shows the trap oscillation frequency $\omega/(2\pi)$ as a function of $I_{00}$. 
The trapping force is much weaker than the damping forces or local dipole forces, so special care must be taken to resolve its contribution. The procedure is as follows. We apply a homogeneous magnetic field along $(\boldsymbol{\hat{x}}+\boldsymbol{\hat{y}})/\sqrt{2}$ for a particular $I_{00}$. We fix the speed at $v=3$ m/s and we solve the OBEs repeatedly for a random choice of directions and for a range of positive and negative magnetic fields. We calculate the component of the force along $(\boldsymbol{\hat{x}}+\boldsymbol{\hat{y}})/\sqrt{2}$ and average together the results from many simulations. For each choice of direction, we ensure that the exact opposite direction is included in the set of simulations. This ensures that the dominant velocity-dependent part of the force (ideally) averages to zero. Because, in the MOT, the magnetic field is proportional to the displacement, the slope of the acceleration at zero magnetic field is proportional to $\omega^2$. The uncertainty in determining $\omega$ is large, because of the noise from the residual damping force. Nevertheless, the oscillation frequencies agree reasonably well with the experimentally measured results. In the CaF MOT, the confining force is expected to arise primarily from a dual-frequency effect between the two laser components closest in frequency to the $F=2$ hyperfine component~\cite{Tarbutt2015b}. This dual-frequency effect should only occur if the two frequency components have opposite polarizations. We have simulated the MOT trapping force for the case where all frequency components have identical polarization, and find that the trapping force vanishes. This result verifies that the dual-frequency mechanism is indeed the mechanism responsible for trapping the molecules. 

The capture velocity of the MOT can be found by calculating the maximum velocity a molecule can have as it enters the MOT region (taken to be at a radius of 10~mm) if it is to be captured. Using the force map $\tilde{F}(\boldsymbol{x},v)$ for the radial plane, at an intensity of $I_{00}=468$ mW/cm$^2$, we calculate a capture velocity of 14~m/s. This is close to the measured value of {$11.2^{+1.2}_{-2.0}$ m/s}~\cite{Williams2017}.

\section{Summary and conclusions}

We have presented a general method for modeling laser cooling and magneto-optical trapping of atoms and molecules, and used the method to understand recent results from experiments with CaF molecules. Our method uses generalized optical Bloch equations to calculate the three-dimensional steady-state force and momentum diffusion constant, taking into account all relevant levels of the molecule and all frequency components of the light. Then, we use the Fokker-Planck-Kramers equation to determine the evolution of the velocity distribution, and the steady-state distribution. Our simulation results show broad agreement with experimental results across a wide range of parameters, and help to improve our understanding of laser cooling and magneto-optical trapping of molecules.

In our previous work~\cite{Devlin2016}, we considered model systems with just one hyperfine ground state and one hyperfine excited state. We found that, for type-II systems driven by red-detuned light, there is Doppler cooling at high speed, but Sisyphus heating at low speed. For blue-detuned light the forces are reversed. In the present work, we have modeled all the levels of CaF relevant to laser cooling, and find that the velocity-dependent force curves are very similar to those found for the simpler systems. Notably, the force curve crosses zero at two speeds: zero, and a specific speed where the Doppler and Sisyphus forces cancel. By solving the Fokker-Planck-Kramers equation, we find the velocity distributions resulting from these unusual force curves. When the light is blue-detuned, and the initial velocity is small enough, the velocity is damped towards zero and the distribution is approximately thermal. In this regime, the temperature can be determined from the damping constant at low velocity, and the momentum diffusion constant. When the light is red-detuned, the velocity distribution is far from being a thermal distribution, and peaks near the special velocity where the force crosses zero, as we would intuitively expect. Despite its non-thermal nature, we find that the ballistic expansion of a cloud with such a velocity distribution is similar to that of a Maxwell-Boltzmann distribution, allowing a temperature to be assigned. This temperature is predicted to be about 10~mK at the highest intensities used in the experiments, and is predicted to decrease as the laser intensity is lowered, mainly because the zero crossing of the force curve shifts to lower speed at lower intensity.  All these predictions agree well with experimental results. Thus, we quantitatively confirm the hypothesis that the balance between Doppler cooling and Sisyphus heating is responsible for the high temperatures observed in molecular MOTs, and also in type-II atomic MOTs. In this regime, momentum diffusion does not play a strong role in determining the temperature. 

The excited-state population calculated from the optical Bloch equations is smaller than predicted by a rate model~\cite{Tarbutt2015,Tarbutt2015b} at all speeds, and drops dramatically at low speeds as the molecules are optically pumped into transient dark states. We find that dark states formed between Zeeman sub-levels of a particular hyperfine level, and Raman dark states which are superpositions of Zeeman sub-levels from different hyperfine states, all play a role in reducing the scattering rate. Similarly, non-adiabatic transitions out of these dark states are important to the Sisyphus cooling / heating mechanism in this system. Our calculated scattering rate in the MOT agrees very well with the measured values across a wide range of intensities. We note that the excited-state population does not drop to zero at zero velocity, showing that there are no time-independent dark states. This is because different frequency components of the light have different polarizations. When a type-II transition is driven by two components of light, there is a time-independent dark state if they have the same polarization (but different detunings), or if they have the same detuning (but different polarizations). If the different frequency components have different polarizations and different detunings, we find there is no time-independent dark state, except in some special cases.

Previous work using a rate model approach concluded that the confining force in a static MOT of CaF is due mainly to a dual-frequency mechanism that arises when two frequency components of opposite polarisation address the same transition with different detunings~\cite{Tarbutt2015b}. This mechanism, which is also analyzed in Ref.~\cite{Cournol2016}, provides both Doppler cooling and strong confinement in cases where little or no confining force is present with one frequency component alone, or with two components of the same polarization. Our simulations using the OBEs support the conclusion that this dual-frequency mechanism is responsible for the trapping force in the CaF MOT. 

Simulations of molecules loaded from a MOT into a molasses predict that the temperature drops on a timescale of about 100~$\mu$s, similar to what is measured. Heating in the molasses is due to the randomness of photon absorption and spontaneous emission events, and due to fluctuations of the dipole force. Our model does not include the last of these, which is particularly difficult to calculate. Across a wide range of parameters, the model consistently predicts steady-state temperatures 3-6 times lower than measured, indicating that the dipole force fluctuations contribute significantly to heating of the molasses. A method for treating this heating mechanism for a multi-level system in 3D would be valuable. At low velocities, the acceleration curve is independent of intensity, but the scattering rate decreases with decreasing intensity. As a result, the temperature of the molasses decreases with decreasing intensity. This is seen in the simulations and the experiments. The cooling time gets longer at low intensities however, and at very low intensities there seems to be no cooling at all. Applying a magnetic field to the molasses increases the scattering rate at zero velocity, and decreases the damping constant, so the temperature increases with magnetic field. The model predicts a quadratic dependence of the temperature on magnetic field, which is also the dependence seen experimentally.

The methods presented in this work can also be used to study laser cooling and trapping of other diatomic or polyatomic species, or for investigating unusual magneto-optical trapping arrangements. To this end, we have presented the equations in a general form so that they can be used by others. We have already used our model to study a blue-detuned MOT of $^{87}$Rb~\cite{Jarvis2018}, and to investigate laser cooling of SrF and YbF molecules~\cite{Lim2018}, where we find the same qualitative behavior as described here for CaF. Future applications include the study of MOTs for molecules with very different energy level structures, the investigation of $\Lambda$-enhanced gray molasses cooling~\cite{Cheuk2018} and other novel cooling schemes, and the study of laser cooling within optical dipole traps~\cite{Anderegg2018}.

Underlying data may be accessed from Zenodo\footnote{\lowercase{10.5281/zenodo.1473592}} and used under the Creative Commons CCZero license.

\acknowledgements
This research was supported by STFC grant no. ST/N000242 and by EPSRC under grants EP/P01058X/1 and EP/M027716/1. We are grateful to the CaF team at Imperial College London for helpful discussions and for providing experimental data.

\bibliography{referencesShort,references}

\end{document}